\documentclass[sigconf]{acmart}

\usepackage{enumitem}
\usepackage{subcaption}
\usepackage{multirow}
\usepackage{colortbl}
\usepackage[normalem]{ulem}
\useunder{\uline}{\ul}{}

\definecolor{bgc}{HTML}{DBE7FA}

\newcommand{\Uset}{\mathcal{U}}
\newcommand{\Iset}{\mathcal{I}}
\newcommand{\Wset}{\mathcal{W}}
\newcommand{\Dset}{\mathcal{D}}

\newcommand{\Hset}{\mathcal{H}}

\newcommand{\linearx}[3]{{#3}\textbf{W}^{#2}_{#1}+\textbf{b}^{#2}_{#1}}
\newcommand{\real}[2]{\mathbb{R}^{#1\times#2}}
\newcommand{\realx}[1]{\mathbb{R}^{#1}}

\AtBeginDocument{%
  }

\setcopyright{acmlicensed}
\copyrightyear{2018}
\acmYear{2018}
\acmDOI{XXXXXXX.XXXXXXX}

\acmConference[Conference acronym 'XX]{Make sure to enter the correct
  conference title from your rights confirmation email}{June 03--05,
  2018}{Woodstock, NY}
\acmISBN{978-1-4503-XXXX-X/18/06}

\begin{document}


\title{Counterfactual Learning-Driven Representation Disentanglement for Search-Enhanced Recommendation}

\renewcommand{\shorttitle}{Counterfactual Learning-Driven Representation Disentanglement for Search-Enhanced Recommendation}
\author{Jiajun Cui\textsuperscript{1}, Xu Chen\textsuperscript{2}, Shuai Xiao\textsuperscript{2}, Chen Ju\textsuperscript{2}, Jinsong Lan\textsuperscript{2}, Qingwen Liu\textsuperscript{2} and Wei Zhang\textsuperscript{1}}
\authornote{Corresponding author.}
\affiliation{
    \institution{\textsuperscript{1}East China Normal University, \textsuperscript{2}Alibaba Group}
    \country{China}
}

\begin{abstract}
For recommender systems in internet platforms, search activities provide additional insights into user interest through query-click interactions with items, and are thus widely used for enhancing personalized recommendation.
However, these interacted items not only have transferable features matching users' interest helpful for the recommendation domain, but also have features related to users' unique intents in the search domain.
Such domain gap of item features is neglected by most current search-enhanced recommendation methods.
They directly incorporate these search behaviors into recommendation, and thus introduce partial negative transfer.
Tackling this problem is challenging due to the lack of explicit supervision signals to disentangle search interactions' different features matching search-specific intent or general interest.
To address this, we propose ClardRec, a \textbf{C}ounterfactual \textbf{l}e\textbf{a}rning-driven \textbf{r}epresentation \textbf{d}isentanglement framework for search-enhanced recommendation, based on the common belief that a user would click an item under a query not solely because of the item-query match but also due to the item's query-independent general features (e.g., color or style) that interest the user.
These general features exclude the reflection of search-specific intents contained in queries, ensuring a pure match to users' underlying interest to complement recommendation.
According to counterfactual thinking, \textit{how would user preferences and query match change for items if we removed their query-related features in search,} we leverage search queries to construct counterfactual signals to disentangle item representations, isolating only query-independent general features.
These representations subsequently enable feature augmentation and data augmentation for the recommendation scenario.
Comprehensive experiments on real datasets demonstrate ClardRec is effective in both collaborative filtering and sequential recommendation scenarios.
\end{abstract}



\maketitle

\begin{figure*}[!t]
\centering
\begin{minipage}{0.29\linewidth}
\includegraphics[width=\linewidth]{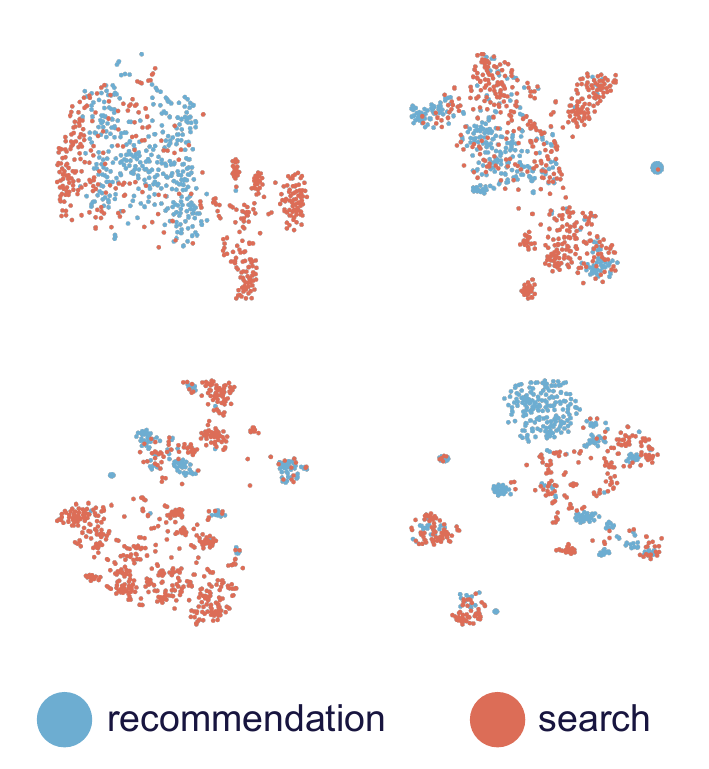}
\subcaption{}
\label{fig:intro_scatter}
\end{minipage}
\begin{minipage}{0.29\linewidth}
\includegraphics[width=\linewidth]{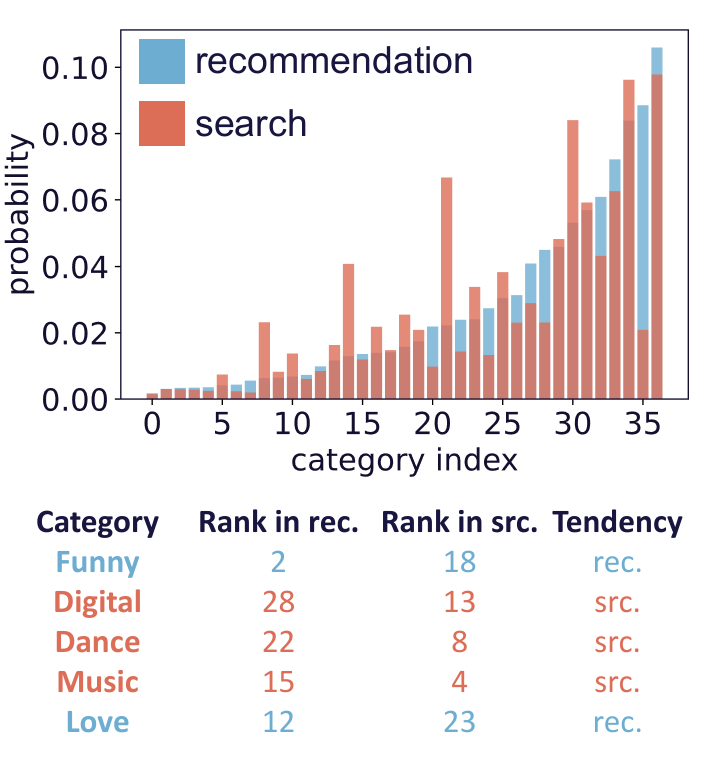}
\subcaption{}
\label{fig:intro_bar}
\end{minipage}
\begin{minipage}{0.29\linewidth}
\includegraphics[width=\linewidth]{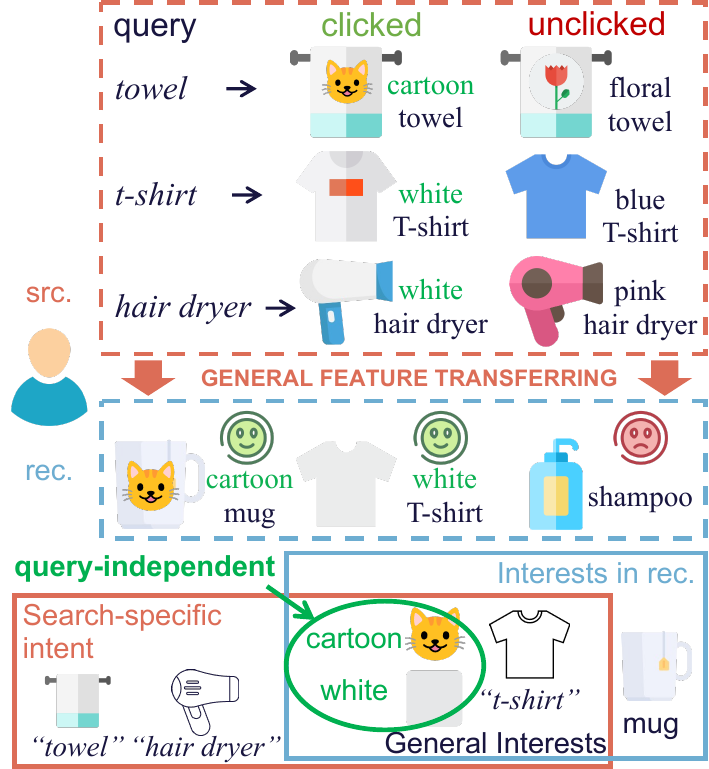}
\subcaption{}
\label{fig:intro_case}
\end{minipage}
\vspace{-.8em}
\caption{
(a) t-SNE visualization of four active users' different clicked product distributions in search and recommendation.
Each product is represented by the average of its title word embeddings derived from~\cite{song2018}.
The active users are sampled from users with more than 800 product click behaviors in a week.
(b) The upper part shows the different probability distributions of clicked videos' 36 main categories in search and recommendation.
The index order of the categories is sorted by their click probabilities in recommendation.
The lower part demonstrates five categories with most probability rank gap between search and recommendation. ``src.'' and ``rec.'' denote search and recommendation.
``Tendency'' denotes which domain the category tends to be clicked.
(c) A toy example of transferring one user's preferred general features from search to recommendation.
The green circled product features are query-independent and satisfy the user's general interest helpful for recommendation.
}
\label{fig:intro}
\vspace{-.8em}
\end{figure*}

\section{introduction}

In recent decades, the information explosion has spurred the emergence of large-scale Internet platforms, such as e-commerce and short video platforms.
These platforms incorporate recommender systems (RS) that suggest personalized items (e.g., products or short videos) to meet users' individual needs.
The internal models of these systems are continuously updated with abundant user behavior data.
Among these behaviors, not only user interactions within the original recommendation scenarios but also their search behaviors provide rich insights into their needs.
The clicked items under search queries have informative features indicating users' underlying interest in recommendation as well. 
Consequently, many researchers have focused on this area, proposing search-enhanced recommendation approaches~\cite{zamani2018,yao2021,si2023_1,si2023_2}.

However, most of these methods directly model and transfer clicked items in the search domain to the recommendation domain, overlooking the distinction of two types of item features in search:
Some match users' general interest that are helpful to compliment recommendation, but others only reflect users' search-specific intent.
This feature gap is evident not only in the behaviors of individual users but also across the overall user group.
For individual users, we conducted a data analysis that visualizes the different distributions of clicked products in search and recommendation for several active users from an e-commerce platform.
As illustrated in Figure~\ref{fig:intro_scatter}, there is a clear discrepancy between the two distributions for all these users, suggesting that users have unique item feature preferences for search and recommendation scenarios, in addition to their overlapping preferences.
For the overall user group, we also analyzed this phenomenon with another public search and recommendation short-video dataset KuaiSAR~\cite{sun2023}, for not losing generality.
As shown in Figure~\ref{fig:intro_bar}, the probability distribution gap of clicked videos' main categories indicates different video feature preferences of users in these two domains.
This domain gap is mainly caused by the different user needs in search and domain.
Users in recommendation usually focus on items satisfying their passive and exploratory interest, such as the illustrated video categories \textit{funny} and \textit{love}.
Contrarily, item features in search tend to reflect users' purposeful and proactive intent, such as actively seeking specific content related to \textit{dance}, \textit{music}, or \textit{digital}.
These differences suggest that directly transferring item features in search to enhance recommendation models could lead to a negative transfer issue due to the domain gap.
Certain item features are specific to search and not beneficial for recommendation, and thus should be excluded from the features that satisfy users' general interest.

Although crucial in search-enhanced recommendation, the lack of explicit supervision makes it challenging to determine whether clicked items' features in search align with user search-specific intents or general interest helpful for the recommendation scenario.
Some works~\cite{si2023_2, shi2024} leverage contrastive learning to uncover shared interest across users' item interaction sequences in both scenarios.
However, this self-supervised training scheme utilizes recommendation behaviors to identify shared information, by which the recommendation does not explicitly extract information from the search domain.
To address this issue, we investigate the underlying correlation among users, items, and queries, based on a widely held belief in information retrieval: \textit{a user's click on an item in response to a specific query is influenced not just by the relevance between the item and the query, but also by the item's general features that appeal to the user, regardless of the query}~\cite{shen2022,cai2014,Micarelli2007}.
Figure~\ref{fig:intro_case} presents a toy example of online shopping to illustrate this belief.
Given the user's search behaviors within the orange dashed box, we cannot determine if the query-matched product features (\textit{towel}, \textit{T-shirt}, and \textit{hair dryer}) correspond to the user's search-specific intents or general interest.
However, the query-independent features (\textit{cartoon} and \textit{white}) reflect the general characteristics of the items, such as colors or styles.
They satisfy the user's general interest, because these features are unrelated to the issued queries yet still preferred by the user.
The blue dashed box illustrates a case of the user's underlying interest in recommendation.
The \textit{mug} interests the user in recommendation, the \textit{T-shirt} is preferred in both scenarios, and the \textit{shampoo} is unlikable despite its relevance to the user's search-specific intents (\textit{towel} and \textit{hair dryer}).
Discriminating general interest and search-specific intents lacks explicit supervision signals.
However, after transferring the general features from the search domain, the user's final interest of a \textit{cartoon mug} and a \textit{white T-shirt} could be mined.
Therefore, these query-independent general item features can be leveraged to enhance recommendations.

Based on this, we propose a \textbf{C}ounterfactual \textbf{l}e\textbf{a}rning-driven \textbf{r}epresentation \textbf{d}isentanglement framework for search-enhanced recommendation (ClardRec), which leverages counterfactual signals to supervise the query-independent item representation learning. 
In particular, we design the counterfactual signals by answering the following question: \textit{How would the query-item and user-item matching score change if we removed the query-related and query-independent features of item representations?}
Applying counterfactual thinking, when the query-related features of item representations are removed, the query-item matching scores will change more significantly than the user-item preference scores.
Conversely, when the query-independent features of item representations are removed, the user-item preference scores will change more significantly due to the removal of user general interest.
We investigate such relationships among users, queries, and items, and propose triplet counterfactual objectives to explicitly supervise the learning of query-independent and query-related item representations.

Furthermore, upon acquiring query-independent item representations from search, we exploit two strategies to jointly facilitate the knowledge transfer to the recommendation domain in an end-to-end manner. 
(i) \textbf{Feature augmentation}:
In the recommendation domain, we use a gated fusion network to augment the representation of each candidate item by integrating its disentangled query-independent representation from the search domain.
The fused item representation, now enriched with additional features, is used to calculate the user's final preference score for the item.
(ii) \textbf{Data augmentation}:
Since the clicked item contains more query-independent features that satisfy the user in the search domain, we regard it as a more considerable positive signal in recommendation.

To the best of our knowledge, this work pioneers the leverage of counterfactual learning to enhance recommendation by transferring query-independent item general features from search.
The main contributions of this paper are as follows:

\begin{itemize}[leftmargin=*]
\item
We identify the domain gap of item features reflecting user search-specific intent and general interest between search and recommendation.
Based on this, we reveal that current search-enhanced recommendation methods lack supervision signals to discriminate such intent and interest, causing a negative transfer issue.
This issue is also confirmed in our conducted experiments.
\item
To address the negative transfer issue, we first discover that query-independent item features can reflect users' general interests.
We then employ counterfactual thinking and propose ClardRec, which constructs supervision signals to disentangle items' query-independent features.
These features subsequently enable knowledge transfer by two strategies, feature augmentation and data augmentation for the recommendation scenario.
\item
Comprehensive experimental results show that ClardRec, adapting to four backbones, exhibits superior performance in collaborative filtering and sequential recommendation on real datasets.

\end{itemize}

\section{related work}

\subsection{Search-enhanced Recommendation}

In recommendation, numerous studies have shown that cross-domain features could benefit every single domain~\cite{chen2023,chen2024,zhu2021,zhu2019}.
This is also applicable for search-enhanced recommendation due to the underlying user interest reflected in user query-click behaviors in search.
Many researchers have devoted efforts to this area.
Some works regard this as a joint optimization problem and apply joint learning to benefit both search and recommendation.
Zamani et al.~\cite{zamani2018} presented the first work that directly uses shared lower-level embeddings and domain-specific upper-level towers to jointly train both search and recommendation models.
Yao et al.~\cite{yao2021} unified behaviors in both domains by assuming that a clicked item in recommendation could be seen as resulting from an empty search query.
Although such unified approaches optimize both search and recommendation models in a general way, they often overlook the extraction of informative domain-specific features.

Another type of methods focuses on recommendations, using search behaviors as auxiliary information. Si et al.~\cite{si2022, si2023_1} embedded search behaviors as instrumental variables to help decompose item embeddings in recommendation to purify the causal part of users' intent.
Subsequently, the study~\cite{si2023_2} used contrastive learning to separate similar and dissimilar interests across search and recommendation in a self-supervised manner.
Our proposed ClardRec belongs to this category but involves novel search auxiliary tasks to provide valuable signals for learning query-independent item general features, which reflect users' interest helpful for recommendation.
This scheme extracts users' general interests from search behaviors mixing search-specific intent, which is overlooked by current search-enhanced recommendation methods. 

\subsection{Counterfactual Learning}

Counterfactual learning has been widely used in various fields to address problems of model interpretability, bias, and data augmentation. For model interpretability, this technique detects the most influential historical interactions of users that impact their current preferences. For example, in a previous study~\cite{tran2021}, the impact of each item on the model’s prediction is measured, and counterfactual explanations are generated. Similar counterfactual thinking is widely used in other fields~\cite{goyal2019,wang2020,zemni2023,yang2020,cui2024interpretable}.
For model debiasing, counterfactual learning has shown efficiency, especially for recommender systems. For example, the previous works \cite{wei2021,zhang2021} address the popularity bias in recommender systems by performing counterfactual inference to remove or even utilize the effect of item popularity.
Some works employ counterfactual learning to construct causal embeddings to improve model unbiasedness~\cite{jian2021,bonner2018}.
Other fields also benefit from this technique to eliminate specific model or data biases~\cite{qian2021,Abbasnejad2020,Parvaneh2020}.
For data augumentation, Ji et al.~\cite{ji2023} iteratively generated hard sequential examples for the model to learn from, boosting performance using counterfactual and logical reasoning. Similarly, Wang et al.~\cite{wang2021} augmented sequence samples by applying counterfactual learning in the item sequence hidden space. Our proposed ClardRec belongs to this category but specifically uses counterfactual learning to disentangle item general features from search to augment recommendation.

\begin{figure*}[!t]
\centering
\includegraphics[width=0.91\linewidth]{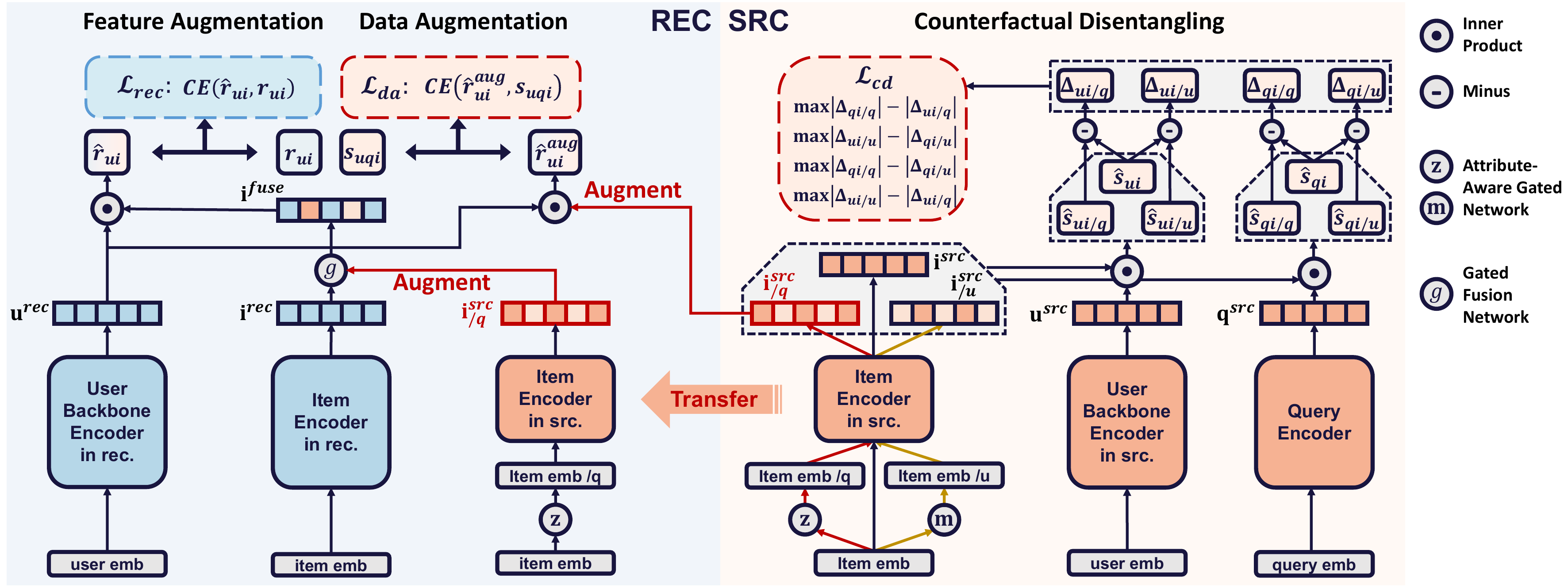}
\vspace{-.5em}
\caption{
Overview of ClardRec. ``REC'' and ``SRC'' denote the recommendation and search domains. The input of user historical sequences to the user backbone encoders in the sequential scenario is omitted for brevity.
}
\label{fig:overview}
\vspace{-.8em}
\end{figure*}

\section{Preliminary}
This section briefly formulates the search-enhanced recommendation task, involving collaborative filtering and sequential recommendation.
Suppose we have a user set $\Uset$, an item set $\Iset$, and a word set $\Wset$ containing all possible words in search queries, then:

(1) For collaborative filtering, we have user interaction data $\Dset_{rec}=\{(u,i) | r_{ui}=1, u \in \Uset, i \in \Iset\}$, where $r_{ui}=1$ indicates that user $u$ has interacted with item $i$ in the recommendation context. 
Besides, there is an interaction set from the search domain, $\Dset_{src}=\{(u,i,q) | s_{uqi}=1, u \in \Uset, i \in \Iset\}$, where $s_{uqi}=1$ indicates that user $u$ has interacted with item $i$ under the search query $q$ with words $q=\{w_1, \ldots, w_{|q|} | w \in \Wset\}$.
The search-enhanced general recommendation aims to train a model $R(\cdot)$ that estimates each user $u$'s underlying preference score $\hat{r}_{ui} = R(u,i)$ for item $i$, given data $\Dset_{rec}$ and $\Dset_{src}$.

(2) For sequential recommendation, each user's interaction data is given as a sequence recording the historical item clicks in search and recommendation, i.e., $\Hset^u_{rec}=\{i^{rec}_1, \ldots, i^{rec}_{|\Hset^u_{rec}|}\}$ and $\Hset^u_{src}=\{(q_1, i^{src}_1), \ldots, (q_{|\Hset^u_{src}|}, i^{src}_{|\Hset^u_{src}|})\}$, where $i^{rec}_t$ is the user's $t$-th clicked item in recommendation, and $i^{src}_t$ is the user's $t$-th clicked item with the issued query $q_t$ in search. Given user $u$'s histories $\Hset^u_{rec}$ and $\Hset^u_{src}$ in the two domains, the search-enhanced sequential recommendation aims to estimate the user's preference score $\hat{r}_{ui} = R(u,i,\Hset^u_{rec},\Hset^u_{src})$ for the next item $i$.
It is worth noting that users and items have informative attributes, such as gender or price. These attributes are omitted in the task formulation for brevity.

\section{Methodology}
Figure~\ref{fig:overview} shows ClardRec, consisting of three main modules: (1) \textbf{Counterfactual Disentangling} that decouples query-independent item features from search, (2) \textbf{Feature Augmentation} that uses decoupled features to complement item features in recommendation, and (3) \textbf{Data Augmentation} that augments decoupled item features in search as users' general interest for recommendation.
In this section, we first introduce the model input and output schemes of ClardRec and then elaborate on these three modules in order.

\subsection{Model Input \& Output}

Due to the overlap of users and items in the search and recommendation domains, we represent them using shared embeddings in the bottom input layers, inspired by search-enhanced recommendation methods with joint-learning schemes~\cite{zamani2018,yao2021}.
For those scenarios with no overlap, we consider it as our future research, as discussed in Appendix~\ref{app:limitation}.
Each user $u$ and item $i$ has a unique ID, $u^{\text{ID}}$ and $i^{\text{ID}}$, with various attributes $u^a = \{u^{a1}, u^{a2}, \ldots\}$ and $i^a = \{i^{a1}, i^{a2}, \ldots\}$.
The number of attributes depends on the data source.
Then, the integrated user embedding is the concatenation of the ID and attribute embeddings:
\begin{equation}
\textbf{u}^{emb} = [\textbf{u}^{\text{ID}}| \textbf{u}^{a1}| \textbf{u}^{a2}|\cdots],
\label{eq:user_emb}
\end{equation}
where $\textbf{u}^{\text{ID}}$ and $\{\textbf{u}^{a1}, \textbf{u}^{a2}, \ldots\}$ denote the user ID and attribute embeddings with dimension size $d_e$. The notation $\mid$ denotes the concatenation operation.
Similarly, we derive the integrated item embedding $\textbf{i}^{emb} = [\textbf{i}^{\text{ID}}|\textbf{i}^{a1}|\textbf{i}^{a2}|\ldots]$.
The obtained user embeddings are fed into subsequent collaborative filtering or sequential recommendation backbone encoders to output their representations, full of underlying comprehensive features, formulated as:
\begin{equation}
\textbf{u}^{rec} = 
\begin{cases}
    \textbf{Enc}^u_{rec}(\textbf{u}^{emb}),\quad &\text{if collaborative filtering},\\
    \textbf{Enc}^u_{rec}(\textbf{u}^{emb}, \textbf{H}^u_{rec}),\quad &\text{if sequential},\\
\end{cases}
\label{eq:user_enc_rec}
\end{equation}
where $\textbf{H}^u_{rec}$ is the embedding set of user $u$'s historical interacted item sequence set $\Hset^u_{rec}$ in recommendation.
It is worth noting that this user encoder $\textbf{Enc}^u_{rec}$ could be adaptive to multiple recommendation backbones for different tasks, which is practiced in subsequent experiments.
Details of the employed backbones could be referred to in Appendix~\ref{app:backbone}.
For items, we use an item encoder to derive their representations as $\textbf{i}^{rec} = \textbf{Enc}^i_{rec}(\textbf{i}^{emb})$.
In practice, the item encoder is designed as a multi-layer perceptron (MLP).
For straightforward recommendation, the preference score of user $u$ for item $i$ is calculated by the inner product of their representations:
\begin{equation}
\hat{r}_{ui}=\textbf{u}^{rec}\cdot\textbf{i}^{rec}.
\label{eq:rec_score}
\end{equation}
This score is then used to optimize the model by a cross-entropy (CE) loss with those preference scores calculated for $M$ randomly selected negative item samples denoted as set $\mathcal{I}^-=\{i^-|r_{ui^-}=0\}$:
\begin{equation}
CE(\hat{r}_{ui},r_{ui})=-\frac{1}{1+M}\sum_{i'\in\{i\}\cup\mathcal{I}^-}r_{ui'}\log\frac{e^{\hat{r}_{ui'}}}{e^{\hat{r}_{ui}} + \sum_{i^-}e^{\hat{r}_{ui^-}}}.
\label{eq:ce}
\end{equation}
Here, we omit the averaging operation over the dataset for brevity.

Besides, ClardRec aims to leverage search information to enhance recommendation.
We propose another set of user backbone encoders and item encoders to derive their representations in search:
\begin{equation}
\textbf{u}^{src} = 
\begin{cases}
    \textbf{Enc}^u_{src}(\textbf{u}^{emb}),\quad &\text{if collaborative filtering},\\
    \textbf{Enc}^u_{src}(\textbf{u}^{emb}, \textbf{H}^u_{src}),\quad &\text{if sequential},\\
\end{cases}
\label{eq:user_enc_src}
\end{equation}
and $\textbf{i}^{src} = \textbf{Enc}^i_{src}(\textbf{i}^{emb})$.
For the query encoder, we use the average pooling of the word embeddings of each query as their representations: $\textbf{q}^{src} = \sum_{w \in q} \textbf{w} / |q|$.
Note that the proposed framework is flexible for different query encoding strategies and we adopt this strategy for its simplicity.
Besides, the user and item encoders are domain-specific and thus do not share parameters between search and recommendation.
Through these encoders, all these user, item, and query representations have the same hidden dimension size $d_h$ and are utilized for the subsequent modules.

\subsection{Counterfactual Disentangling}

Based on the assumption that query-independent features in search reflect user interest and are thus transferable to recommendation, we propose counterfactual disentangling during the item encoding.

\subsubsection{Item Attribute-aware Disentangling}

This module employs an item attribute-aware gated network to linearly decompose each item embedding into two parts, $\textbf{i}^{emb}_{/q}$ and $\textbf{i}^{emb}_{/u}$, representing the item embedding with and without query-related features.
Among them, $\textbf{i}^{emb}_{/q}$ is the embedding without features for queries and
$\textbf{i}^{emb}_{/u}$ contains features only for queries, neglecting the general features interesting to users.
We thus use the notation “$/u$” for clarity in the subsequent description.
According to this, the decomposition is derived from an item attribute-aware gated network that assigns different weights for the concatenation of the two disentangled embeddings.
This process is formulated as follows:
\begin{equation}
    \begin{aligned}
    &\textbf{i}^{emb}_{/q} = 2\cdot[\textbf{z}^{\text{ID}}\otimes\textbf{i}^{\text{ID}}| z^{a1}\cdot\textbf{i}^{a1}| z^{a2}\cdot\textbf{i}^{a2}|\cdots],\\
    &\textbf{i}^{emb}_{/u} = 2\cdot[\textbf{m}^{\text{ID}}\otimes\textbf{i}^{\text{ID}}| m^{a1}\cdot\textbf{i}^{a1}| m^{a2}\cdot\textbf{i}^{a2}|\cdots],
\end{aligned}
\label{eq:attribute_gate}
\end{equation}
where $\otimes$ denotes the Hadamard product. 
$\textbf{z}^{ID}\in\realx{d_e}$ is the element-wise gate control vector of the ID attributes for the item embedding without query-related features, generated by an MLP:
\begin{equation}
\textbf{z}^{\text{ID}} = \sigma\left(\linearx{ID}{2}{\text{ReLU}\left(\linearx{ID}{1}{\textbf{i}^{emb}}\right)}\right)
\label{eq:ID_gate}
\end{equation}
and $z^{aj}\in\mathbb{R}$ is the $j$-th value of the attribute-specific gate control vector $\textbf{z}^a\in\realx{|i^a|}$, which is generated by another MLP:
\begin{equation}
\textbf{z}^{a} = \sigma\left(\linearx{a}{2}{\text{ReLU}\left(\linearx{a}{1}{\textbf{i}^{emb}}\right)}\right).
\label{eq:attr_gate}
\end{equation}
Among these, $\textbf{W}^{1}_{ID}\in\real{d_i}{d_h},\textbf{W}^{2}_{ID}\in\real{d_h}{d_e},\textbf{b}^{1}_{ID}\in\realx{d_h},\textbf{b}^{2}_{ID}\in\realx{d_e},\textbf{W}^{1}_{a}\in\real{d_i}{d_h},\textbf{W}^{2}_{a}\in\real{d_h}{
|i^a|},\textbf{b}^{1}_{a}\in\realx{d_h},\textbf{b}^{2}_{a}\in\realx{|i^a|}$ are the learnable parameters in the MLPs.
$d_i = (|i^a| + 1)\cdot d_e$ is the number of item embedding dimensions and $d_h$ is the number of hidden dimensions.
$\text{ReLU}(\cdot)$ and $\sigma(\cdot)$ are the activation and sigmoid functions.
The gate values for the item embedding containing only features for queries are calculated by $\textbf{m}^{ID} = \textbf{1} - \textbf{z}^{ID}$ and $m^{aj} = 1 - z^{aj}$.

Based on this, the two generated embeddings from the gates could satisfy the constrain $\textbf{i}^{emb} = (\textbf{i}^{emb}_{/q} + \textbf{i}^{emb}_{/u})/2$, formulating a disentanglement of the item embedding.
This gated control scheme captures the fine-grained differences at the item attribute level, as different attributes may have varying tendencies for query matching or user general interest. For example, the item category ``digital'' is more frequently searched for query matching than ``love''.
After decomposing the embeddings, the two separated embeddings are fed into the item encoder to generate two comprehensive disentangled item representations, with or without features for queries:
\begin{equation}
\textbf{i}^{src}_{/q} = \textbf{Enc}^i_{src}(\textbf{i}^{emb}_{/q}),\quad\textbf{i}^{src}_{/u} = \textbf{Enc}^i_{src}(\textbf{i}^{emb}_{/u}).
\label{eq:item_wo_encode}
\end{equation}
\subsubsection{Triplet Counterfactual Objective}
The generated item representations with and without features for queries originate from the learnable gated control.
Henceforth, we apply user-query-item triplet counterfactual objectives to learn the disentangled item representations with query-related or query-independent information, alleviating lacking supervision signals for disentanglement. 

\textbf{Motivation:} This approach is based on the counterfactual thinking that if we remove the query-related features of an item, the degree of its match to a query should vary more than the degree of its match to a user's interest.
We denote this as \textbf{matching-wise} counterfactual signals.
Correspondingly, we could transform this statement as:
The degree to which an item matches a query should vary more if we remove its query-related features rather than its query-independent features.
We denote this as \textbf{removal-wise} counterfactual signals.
Based on these two types of signals, we design the triplet counterfactual objective as follows:

(\romannumeral1) \textbf{Matching-wise.} Denote the inner products $\hat{s}_{ui} = \textbf{u}^{src}\cdot\textbf{i}^{src}$ and $\hat{s}_{ui/q} = \textbf{u}^{src}\cdot\textbf{i}^{src}_{/q}$ as the user-item preference scores before and after removing the item query-related features.
Similarly, $\hat{s}_{qi} = \textbf{q}^{src}\cdot\textbf{i}^{src}$ and $\hat{s}_{qi/q} = \textbf{q}^{src}\cdot\textbf{i}^{src}_{/q}$ are the query-item match scores before and after the same removal.
Suppose the variations are defined as $\Delta_{ui/q} = \hat{s}_{ui/q} - \hat{s}_{ui}$ and $\Delta_{qi/q} = \hat{s}_{qi/q} - \hat{s}_{qi}$.
Then we have:
\begin{equation}
|\Delta_{qi/q}| > |\Delta_{ui/q}|.
\label{eq:cf_loss_wo_q}
\end{equation}
Conversely, when only retaining the query-related features of an item, i.e., removing the general features that interest users, the user-item preference score should vary more:
\begin{equation}
|\Delta_{ui/u}| > |\Delta_{qi/u}|,
\label{eq:cf_loss_wo_u}
\end{equation}
where $\hat{s}_{ui/u} = \textbf{u}^{src}\cdot\textbf{i}^{src}_{/u}$, $\hat{s}_{qi/u} = \textbf{q}^{src}\cdot\textbf{i}^{src}_{/u}$, $\Delta_{ui/u} = \hat{s}_{ui/u} - \hat{s}_{ui}$, and $\Delta_{qi/u} = \hat{s}_{qi/u} - \hat{s}_{qi}$.

(\romannumeral2) \textbf{Removal-wise}. The variation in the query-item match should be larger when removing query-related item features rather than retaining them.
Similarly, the variation in the user-item match should be the opposite.
Then there are:
\begin{equation}
|\Delta_{qi/q}| > |\Delta_{qi/u}|,\quad |\Delta_{ui/u}| > |\Delta_{ui/q}|.
\label{eq:cf_loss_qi_ui}
\end{equation}

\textbf{Formulation of Triplet counterfactual objective:} We transfer these user-query-item variation relationships into triplet counterfactual learning objectives that maximize the difference between the score variations.
Consider the match-wise signals in Equation~\ref{eq:cf_loss_wo_q}. 
A preferred (positive) item $i$ for user $u$ under query $q$ should match the query and interest the user. Therefore, when removing any feature from the item representation, the match and preference scores should drop or remain unchanged (i.e., $\Delta_{qi/q}<=0$ and $\Delta_{ui/q}<=0$).
Conversely, for a disliked (negative) item $i^-$ that the user has never interacted with, none of its features should match the query or interest the user.
The corresponding scores are thus expected to increase or remain unchanged after the removal (i.e., $\Delta_{qi^-/q}>=0$ and $\Delta_{ui^-/q}>=0$).
Similar logic could be used for Equation~\ref{eq:cf_loss_wo_u}.
We thus formulate these as log-sigmoid triplet losses for each $(u,q,i)$ triplet and the sampled $M$ negative items $\{i^-\}$:
\begin{equation}
\begin{aligned}
&\mathcal{L}^{/q}_{cd}=-\log\sigma(\Delta_{qi/q}-\Delta_{ui/q})-\frac{1}{M}\sum_{i^-}\log\sigma(\Delta_{ui^-/q}-\Delta_{qi^-/q}),\\
&\mathcal{L}^{/u}_{cd}=-\log\sigma(\Delta_{ui/u}-\Delta_{qi/u})-\frac{1}{M}\sum_{i^-}\log\sigma(\Delta_{qi^-/u}-\Delta_{ui^-/u}).
\label{eq:cf_loss_wo_log}
\end{aligned}
\end{equation}
For the removal-wise counterfactual signals in Equation~\ref{eq:cf_loss_qi_ui}, we similarly have
\begin{equation}
\begin{aligned}
&\mathcal{L}^{qi}_{cd}=-\log\sigma(\Delta_{qi/q}-\Delta_{qi/u})-\frac{1}{M}\sum_{i^-}\log\sigma(\Delta_{qi^-/u}-\Delta_{qi^-/q}),\\
&\mathcal{L}^{ui}_{cd}=-\log\sigma(\Delta_{ui/u}-\Delta_{ui/q})-\frac{1}{M}\sum_{i^-}\log\sigma(\Delta_{ui^-/q}-\Delta_{ui^-/u}).
\end{aligned}
\label{eq:cf_loss_other_log}
\end{equation}
We then obtain the total triplet counterfactual objective as
\begin{equation}
\mathcal{L}_{cd} = \mathcal{L}^{/q}_{cd} + \mathcal{L}^{/u}_{cd} + \mathcal{L}^{qi}_{cd} + \mathcal{L}^{ui}_{cd}.
\label{eq:cf_loss}
\end{equation}
To ensure reasonable score variation after the counterfactual operations, we also employ a loss term to constrain the score drop for positive items and score increase for negative items after the removals.
Taking $\Delta_{ui/q}$ as an example, we have:
\begin{equation}
\mathcal{L}^{ui/q}_{con} = \max(\Delta_{ui/q},0) + \frac{1}{M}\sum_{i^-}\max(-\Delta_{ui^-/q},0).
\label{eq:con_loss_uiwoq}
\end{equation}
The total constraint loss is then:
\begin{equation}
\mathcal{L}_{con} = \mathcal{L}^{ui/q}_{con}+\mathcal{L}^{ui/u}_{con}+\mathcal{L}^{qi/q}_{con}+\mathcal{L}^{qi/u}_{con}.
\label{eq:con_loss}
\end{equation}

Using this paradigm, we can obtain the item representation $\textbf{i}^{src}_{/q}$, which is rich in query-independent general features for subsequent feature augmentation and data augmentation.

\begin{table*}[!t]
\renewcommand{\arraystretch}{0.88}
\setlength{\tabcolsep}{4.0pt}
\centering
\caption{Main experimental results. The best results among ClardRec and other baselines are in bold. The second ones of baselines are in italic. * indicates statistical significance over the best baseline, measured by T-test with p-value $\leq$ 0.05. ``SA'' denotes if the method leverages auxiliary learning tasks from the search domain. ``CF'' denotes the collaborative filtering.}
\vspace{-.5em}
\label{tab:main_exp}
\begin{tabular}{cccc|cccc|cccc}
\hline
\multicolumn{4}{c|}{Dataset} &
  \multicolumn{4}{c|}{KuaiSAR} &
  \multicolumn{4}{c}{E-commerce R\&S} \\ \hline
\multicolumn{1}{c|}{Scenario} &
  \multicolumn{1}{c|}{SA} &
  \multicolumn{1}{c|}{User encoder} &
  Method &
  HR@1 &
  HR@5 &
  NDCG@5 &
  MRR &
  HR@1 &
  HR@5 &
  NDCG@5 &
  MRR \\ \hline
\multicolumn{1}{c|}{\multirow{8}{*}{CF}} &
  \multicolumn{1}{c|}{\multirow{3}{*}{$\times$}} &
  \multicolumn{1}{c|}{\multirow{3}{*}{MLP}} &
  - &
  0.1125 &
  0.2264 &
  0.1724 &
  0.1756 &
  0.1271 &
  0.3550 &
  0.2429 &
  0.2498 \\
\multicolumn{1}{c|}{} &
  \multicolumn{1}{c|}{} &
  \multicolumn{1}{c|}{} &
  +AUG &
  0.1096 &
  0.2180 &
  0.1667 &
  0.1679 &
  0.1004 &
  0.2857 &
  0.1945 &
  0.2040 \\
\multicolumn{1}{c|}{} &
  \multicolumn{1}{c|}{} &
  \multicolumn{1}{c|}{} &
  +IV4REC+ &
  0.1130 &
  0.2247 &
  0.1717 &
  0.1744 &
  0.1270 &
  0.3553 &
  0.2430 &
  0.2500 \\ \cline{2-12} 
\multicolumn{1}{c|}{} &
  \multicolumn{1}{c|}{\multirow{4}{*}{$\checkmark$}} &
  \multicolumn{1}{c|}{\multirow{2}{*}{MLP}} &
  +JSR &
  0.1129 &
  0.2253 &
  0.1736 &
  0.1759 &
  0.1262 &
  0.3536 &
  0.2417 &
  0.2486 \\
\multicolumn{1}{c|}{} &
  \multicolumn{1}{c|}{} &
  \multicolumn{1}{c|}{} &
  +ClardRec &
  0.1153 &
  0.2276 &
  0.1744 &
  0.1779 &
  0.1306 &
  0.3613 &
  0.2479 &
  0.2541 \\ \cline{3-12} 
\multicolumn{1}{c|}{} &
  \multicolumn{1}{c|}{} &
  \multicolumn{1}{c|}{\multirow{2}{*}{MMoE}} &
  +JSR &
  {\ul 0.1152} &
  {\ul 0.2327} &
  {\ul 0.1768} &
  {\ul 0.1817} &
  {\ul 0.1276} &
  {\ul 0.3633} &
  {\ul 0.2473} &
  {\ul 0.2530} \\
\multicolumn{1}{c|}{} &
  \multicolumn{1}{c|}{} &
  \multicolumn{1}{c|}{} &   
  +ClardRec &
  \textbf{0.1165*} &
  \textbf{0.2415*} &
  \textbf{0.1817*} &
  \textbf{0.1886*} &
  \textbf{0.1317*} &
  \textbf{0.3710*} &
  \textbf{0.2534*} &
  \textbf{0.2581*} \\ \cline{2-12} 
\multicolumn{1}{c|}{} &
  \multicolumn{3}{c|}{\textit{Improv.}} &
  \textit{+1.18\%} &
  \textit{+3.81\%} &
  \textit{+2.77\%} &
  \textit{+3.76\%} &
  \textit{+3.21\%} &
  \textit{+2.14\%} &
  \textit{+2.47\%} &
  \textit{+2.02\%} \\ \hline
\multicolumn{1}{c|}{\multirow{12}{*}{Sequential}} &
  \multicolumn{1}{c|}{\multirow{7}{*}{$\times$}} &
  \multicolumn{1}{c|}{\multirow{3}{*}{GRU4Rec}} &
  - &
  0.3242 &
  0.5092 &
  {\ul 0.4234} &
  0.4185 &
  0.2619 &
  0.5741 &
  0.4243 &
  0.4091 \\
\multicolumn{1}{c|}{} &
  \multicolumn{1}{c|}{} &
  \multicolumn{1}{c|}{} &
  +AUG &
  0.2868 &
  0.4589 &
  0.3779 &
  0.3772 &
  0.2564 &
  0.5402 &
  0.4042 &
  0.3924 \\
\multicolumn{1}{c|}{} &
  \multicolumn{1}{c|}{} &
  \multicolumn{1}{c|}{} &
  +IV4REC+ &
  {\ul 0.3260} &
  0.5074 &
  0.4233 &
  {\ul 0.4185} &
  0.2616 &
  0.5788 &
  0.4266 &
  0.4107 \\ \cline{3-12} 
\multicolumn{1}{c|}{} &
  \multicolumn{1}{c|}{} &
  \multicolumn{1}{c|}{\multirow{3}{*}{SASRec}} &
  - &
  0.3196 &
  0.5112 &
  0.4213 &
  0.4146 &
  0.2880 &
  0.6069 &
  0.4546 &
  0.4379 \\
\multicolumn{1}{c|}{} &
  \multicolumn{1}{c|}{} &
  \multicolumn{1}{c|}{} &
  +AUG &
  0.2770 &
  0.4549 &
  0.3718 &
  0.3703 &
  0.2869 &
  0.5786 &
  0.4395 &
  0.4247 \\
\multicolumn{1}{c|}{} &
  \multicolumn{1}{c|}{} &
  \multicolumn{1}{c|}{} &
  +IV4REC+ &
  0.3235 &
  0.5083 &
  0.4223 &
  0.4151 &
  0.2989 &
  {\ul 0.6259} &
  {\ul 0.4698} &
  {\ul 0.4501}
  \\\cline{3-12}
  \multicolumn{1}{c|}{} &
  \multicolumn{1}{c|}{} &
  \multicolumn{1}{c|}{-} &
  \multicolumn{1}{c|}{\multirow{1}{*}{SESREC}} &
  0.3222 &
  0.4611 &
  0.3978 &
  0.3980 &
  0.2733 &
  0.5937 &
  0.4221 &
  0.3972 
  \\\cline{2-12}
\multicolumn{1}{c|}{} &
  \multicolumn{1}{c|}{\multirow{4}{*}{$\checkmark$}} &
  \multicolumn{1}{c|}{\multirow{2}{*}{GRU4Rec}} &
  +JSR &
  0.3226 &
  {\ul 0.5123} &
  0.4238 &
  0.4170 &
  0.2651 &
  0.5796 &
  0.4298 &
  0.4139 \\
\multicolumn{1}{c|}{} &
  \multicolumn{1}{c|}{} &
  \multicolumn{1}{c|}{} &
  +ClardRec &
  \textbf{0.3389*} &
  0.5126 &
  \textbf{0.4326*} &
  \textbf{0.4263*} &
  0.2923 &
  0.6199 &
  0.4635 &
  0.4437 \\ \cline{3-12} 
\multicolumn{1}{c|}{} &
  \multicolumn{1}{c|}{} &
  \multicolumn{1}{c|}{\multirow{2}{*}{SASRec}} &
  +JSR &
  0.3222 &
  0.5113 &
  0.4229 &
  0.4166 &
  {\ul 0.2996} &
  0.6188 &
  0.4665 &
  0.4478 \\
\multicolumn{1}{c|}{} &
  \multicolumn{1}{c|}{} &
  \multicolumn{1}{c|}{} &
  +ClardRec &
  0.3331 &
  \textbf{0.5140} &
  0.4294 &
  0.4238 &
  \textbf{0.3321*} &
  \textbf{0.6561*} &
  \textbf{0.5024*} &
  \textbf{0.4808*} \\ 
  \cline{2-12} 
\multicolumn{1}{c|}{} &
  \multicolumn{3}{c|}{\textit{Improv.}} &
  \textit{+3.94\%} &
  \textit{+0.33\%} &
  \textit{+2.18\%} &
  \textit{+1.85\%} &
  \textit{+10.84\%} &
  \textit{+4.82\%} &
  \textit{+6.93\%} &
  \textit{+6.82\%} \\ \hline
\end{tabular}
\vspace{-.8em}
\end{table*}

\subsection{Feature Augmentation}

ClardRec leverages the query-independent item representation, reflecting general features and derived from the search encoder, to enhance the item representation in recommendation.
This process is employed by a gated fusion network, another MLP that adaptively combines the original and query-independent item representations from the recommendation and search encoders, respectively:
\begin{equation}
\textbf{i}^{fuse} = g\cdot\textbf{i}^{rec} + (1-g)\cdot\textbf{i}^{src}_{/q},
\label{eq:fuse}
\end{equation}
where the combination factor $g$ is a scalar and calculated by the gated fusion network:
\begin{equation}
g = \sigma\left(\linearx{f}{2}{\text{ReLU}\left(\linearx{f}{1}{[\textbf{i}^{rec}|\textbf{i}^{src}_{/q}]}\right)}\right).
\label{eq:fuse_mlp}
\end{equation}
In this network, $\textbf{W}^1_{f}\in\real{2d_h}{d_h},\textbf{W}^2_{f}\in\real{d_h}{1},\textbf{b}^1_{f}\in\realx{d_h}$ and $\textbf{b}^2_{f}\in\realx{1}$ are the learnable matrices and vectors.
The notation $|$ denotes the concatenation operation.
Hereinafter, we rewrite Equation~\ref{eq:rec_score} and derive the preference score in recommendation as:
\begin{equation}
\hat{r}_{ui}=\textbf{u}^{rec}\cdot\textbf{i}^{fuse},
\label{eq:fa_pred}
\end{equation}
for the final prediction and model training.

\subsection{Data Augmentation}

As previously discussed, search behaviors mix the reflection of user search-specific intent and general interest, making them unsuitable for direct transfer to recommendation systems. However, through counterfactual disentangling, we can decompose the query-independent general item features. The degree of match between these features and users reflects their general interest.
For each interaction pair $(u, i)$ in search (i.e., $s_{uqi} = 1$), such interest can be measured by $\hat{s}_{ui/q}$, the query-independent preference score obtained by counterfactual disentangling.
A higher general interest underlying the search interaction indicates a greater likelihood of the user also preferring those features in recommendation. 
Therefore, we regard this as an auxiliary signal to augment recommendation learning.
We consider $\hat{s}_{ui/q}$ as a confidence score and define $\hat{r}^{aug}_{ui} = \textbf{u}^{rec}\cdot\textbf{i}^{src}_{/q}$ as the transferred preference in recommendation with a weight function $\omega(\cdot)$ monotonically consistent with $\hat{s}_{ui/q}$:
\begin{equation}
\mathcal{L}_{da}=\omega(\hat{s}_{ui/q})CE(\hat{r}^{aug}_{ui},s_{uiq}),
\label{eq:da_pred}
\end{equation}
In practice, we choose the in-batch softmax normalization $\omega(\hat{s}_{ui/q}) = \text{softmax}_{(u,i)\in\mathcal{B}}(\hat{s}_{ui/q})$ which performs the best.
Then, ClardRec could benefit recommendation not only from feature augmentation, but also from a data augmentation perspective.

\subsection{Model Training}

ClardRec is designed as an end-to-end model, with a joint-learning scheme for the model training process.
In addition to the counterfactual disentangling and data augmentation objectives, we also inject supervision signals of user-item preference and query-item match in search.
This endows semantic information to the encoded representations of users, items, and queries, making the disentangled item representations meaningful.
Concretely, for each triplet $(u,q,i)$ in search, we optimize the preference and match scores as:
\begin{equation}
\mathcal{L}_{src}=CE(\hat{s}_{ui},s_{uqi}) + CE(\hat{s}_{qi},s_{uqi}).
\label{eq:src_loss}
\end{equation}
Combining these auxiliary losses in search with the main training loss for recommendation $\mathcal{L}
_{rec}=CE(\hat{r}_{ui},r_{ui})$, the total loss is
\begin{equation}
\mathcal{L}=\mathcal{L}_{rec}+\lambda(\mathcal{L}_{src}+\alpha\mathcal{L}_{cd}+\beta\mathcal{L}_{da}+\gamma\mathcal{L}_{con}),
\label{eq:total_loss}
\end{equation}
where $\lambda,\alpha,\beta$ and $\gamma$ are hyper-parameters to balance the loss.
In practice, most of these hyper-parameter values are selected from a small set
\{0.01,0.1,0.2,0.5,1.0\} and they already lead to good performance, eliminating more tedious tuning processes.

\section{Experiments}

In this section, we conduct comprehensive experiments to evaluate ClardRec and answer the following research questions.
\begin{itemize}[leftmargin=0.7cm]
\item[\textbf{Q1:}] Does ClardRec achieve competitive recommendation performance compared to current state-of-the-art search-enhanced recommendation frameworks and methods?
\item[\textbf{Q2:}] What are the roles and impacts of different components of ClardRec on the overall recommendation performance?
\item[\textbf{Q3:}] Could ClardRec derive query-independent item features matching user general interest by counterfactual disentangling?
\end{itemize}
Additionally, due to space constraints, other experiments such as hyper-parameter analysis are included in Appendix~\ref{app:sup_exp}.

\subsection{Experimental Setting}

\subsubsection{Datasets}
We selected two datasets for experiments.
\begin{itemize}[leftmargin=*]
\item \textbf{KuaiSAR}\footnote{\url{https://kuaisar.github.io/}}~\cite{sun2023}: This public dataset was collected from a short-video platform and covers the search and recommendation behaviors of approximately 25,000 users over 19 days in 2023.

\item \textbf{E-commerce R\&S}: This industrial dataset was collected from an online widely-used e-commerce platform.
We gathered all user click behaviors in search and recommendation over two weeks in 2024.
The dataset was split into two subsets: one week for collaborative filtering recommendation and another week for sequential recommendation.
We filtered users and items with fewer than five click interactions in both search and recommendation.

\end{itemize}

The preprocessing approach of the datasets and the statistics after processing are detailed in Appendix~\ref{app:dataset}.

\begin{table*}[!t]
\renewcommand{\arraystretch}{0.88}
\centering
\caption{Results of the ablation experiments.}
\vspace{-.8em}
\label{tab:abl_exp}
\begin{tabular}{ccl|cccc|cccc}
\hline
\multicolumn{3}{c|}{Dataset} &
  \multicolumn{4}{c|}{KuaiSAR} &
  \multicolumn{4}{c}{E-commerce R\&S} \\ \hline
\multicolumn{1}{c|}{Scenario} &
  \multicolumn{1}{c|}{User encoder} &
  \multicolumn{1}{c|}{Method} &
  HR@1 &
  HR@5 &
  NDCG@5 &
  MRR &
  HR@1 &
  HR@5 &
  NDCG@5 &
  MRR \\ \hline
\multicolumn{1}{c|}{\multirow{8}{*}{CF}} &
  \multicolumn{1}{c|}{\multirow{4}{*}{MLP}} &
  ClardRec &
  \textbf{0.1153} &
  \textbf{0.2276} &
  \textbf{0.1744} &
  \textbf{0.1779} &
  \textbf{0.1306} &
  \textbf{0.3613} &
  \textbf{0.2479} &
  \textbf{0.2541} \\ 
\multicolumn{1}{c|}{} &
  \multicolumn{1}{c|}{} &
  -CD &
  0.1144 &
  0.2266 &
  0.1734 &
  0.1762 &
  0.1284 &
  0.3555 &
  0.2439 &
  0.2507 \\
\multicolumn{1}{c|}{} &
  \multicolumn{1}{c|}{} &
  -FA &
  0.1133 &
  0.2269 &
  0.1730 &
  0.1758 &
  0.1282 &
  0.3547 &
  0.2434 &
  0.2502 \\
\multicolumn{1}{c|}{} &
  \multicolumn{1}{c|}{} &
  -DA &
  0.1143 &
  0.2266 &
  0.1733 &
  0.1769 &
  0.1284 &
  0.3598 &
  0.2460 &
  0.2520 \\ \cline{2-11} 
\multicolumn{1}{c|}{} &
  \multicolumn{1}{c|}{\multirow{4}{*}{MMoE}} &
  ClardRec &
  \textbf{0.1165} &
  0.2415 &
  \textbf{0.1817} &
  0.1886 &
  \textbf{0.1317} &
  \textbf{0.3710} &
  \textbf{0.2534} &
  \textbf{0.2581} \\ 
\multicolumn{1}{c|}{} &
  \multicolumn{1}{c|}{} &
  -CD &
  0.1157 &
  0.2365 &
  0.1789 &
  0.1846 &
  0.1305 &
  0.3679 &
  0.2512 &
  0.2560 \\
\multicolumn{1}{c|}{} &
  \multicolumn{1}{c|}{} &
  -FA &
  0.1145 &
  0.2338 &
  0.1769 &
  0.1836 &
  0.1291 &
  0.3671 &
  0.2500 &
  0.2550 \\
\multicolumn{1}{c|}{} &
  \multicolumn{1}{c|}{} &
  -DA &
  0.1163 &
  \textbf{0.2417} &
  0.1816 &
  \textbf{0.1887} &
  0.1296 &
  0.3655 &
  0.2494 &
  0.2551 \\ \hline
\multicolumn{1}{c|}{\multirow{8}{*}{Sequential}} &
  \multicolumn{1}{c|}{\multirow{4}{*}{GRU4Rec}} &
  ClardRec &
  \textbf{0.3389} &
  \textbf{0.5126} &
  \textbf{0.4326} &
  \textbf{0.4263} &
  \textbf{0.2923} &
  \textbf{0.6199} &
  \textbf{0.4635} &
  \textbf{0.4437} \\
\multicolumn{1}{c|}{} &
  \multicolumn{1}{c|}{} &
  -CD &
  0.3351 &
  0.5090 &
  0.4279 &
  0.4219 &
  0.2818 &
  0.6070 &
  0.4516 &
  0.4327 \\
\multicolumn{1}{c|}{} &
  \multicolumn{1}{c|}{} &
  -FA &
  0.3384 &
  0.5067 &
  0.4294 &
  0.4238 &
  0.2709 &
  0.5997 &
  0.4422 &
  0.4237 \\
\multicolumn{1}{c|}{} &
  \multicolumn{1}{c|}{} &
  -DA &
  0.3376 &
  0.5100 &
  0.4311 &
  0.4250 &
  0.2853 &
  0.6101 &
  0.4549 &
  0.4359 \\ \cline{2-11} 
\multicolumn{1}{c|}{} &
  \multicolumn{1}{c|}{\multirow{4}{*}{SASRec}} &
  ClardRec &
  \textbf{0.3331} &
  \textbf{0.5140} &
  \textbf{0.4294} &
  \textbf{0.4238} &
  \textbf{0.3321} &
  \textbf{0.6561} &
  \textbf{0.5024} &
  \textbf{0.4808} \\
\multicolumn{1}{c|}{} &
  \multicolumn{1}{c|}{} &
  -CD &
  0.3191 &
  0.5045 &
  0.4181 &
  0.4114 &
  0.3066 &
  0.6281 &
  0.4752 &
  0.4554 \\
\multicolumn{1}{c|}{} &
  \multicolumn{1}{c|}{} &
  -FA &
  0.3261 &
  0.5123 &
  0.4249 &
  0.4192 &
  0.3031 &
  0.6247 &
  0.4716 &
  0.4520 \\
\multicolumn{1}{c|}{} &
  \multicolumn{1}{c|}{} &
  -DA &
  0.3324 &
  0.5137 &
  0.4291 &
  0.4237 &
  0.3195 &
  0.6444 &
  0.4901 &
  0.4689 \\ \hline
\end{tabular}
\vspace{-.8em}
\end{table*}

\subsubsection{Baselines.}
To demonstrate the superiority of ClardRec enhancing recommendation, we adapt it to four typical backbones, \textbf{MLP}~\cite{Huang2013} and \textbf{MMoE}~\cite{Ma2018} for the collaborative filtering scenario, and \textbf{GRU4Rec}~\cite{Hidasi2016} and \textbf{SASRec}~\cite{Kang2018} for the sequential scenario.
We compare it against several state-of-the-art search-enhanced recommendation methods, encompassing three learning frameworks and one integrated method except for the original backbones. 
\textbf{AUG} directly transfers the click behaviors in search to recommendation.
We also select a joint-learning framework \textbf{JSR}~\cite{zamani2018} and a causal learning framework \textbf{IV4REC+}~\cite{si2023_1} for comparisons.
For the integrated methods, we choose \textbf{SESRec}~\cite{si2023_2} that models the sharing interest across search and recommendation by contrastive learning.

\subsubsection{Evaluation.} Following some previous works~\cite{si2023_2}, we evaluate ClardRec using three standard metrics in recommendation systems, hit ratio (HR), normalized discounted cumulative gain (NDCG), and mean reciprocal rank (MRR).
For HR, we measure the rate of correctly recommended items in the top-1 and 5 lists among the ground-truth and 99 negative items, denoted as HR@1 and HR@5.
Similarly, we also report NGCG@5.
Each experiment is conducted five times with different random seeds to ensure result robustness.
Besides, we halts training when the performance on the validation set does not improve over 10 consecutive epochs.

\subsubsection{Implementation Details}
We employ the Adam optimizer~\cite{kingma2014} for all methods and carefully tune hyper-parameters to achieve optimal performance.
To ensure fairness in comparison, we fix the embedding and hidden dimension sizes at 32 and 128, respectively.
Additionally, we strictly adhere to the hyper-parameter settings detailed in the original papers of all baseline methods.
For ClardRec, specific hyper-parameter details are provided in Appendix~\ref{app:hyper_setting}.

\subsection{Overall Performance (Q1)}

Table~\ref{tab:main_exp} presents the overall performance comparison between ClardRec and other baseline methods.
ClardRec consistently outperforms the other methods across both datasets with improvements from 0.33\% to 10.84\%.
Notably, all backbones demonstrate significant improvements when augmented with ClardRec, underscoring its efficacy.
Conversely, the AUG framework shows degradation across all backbones, highlighting negative transfer issues caused by transferring the search-specific intent.
Comparing the two datasets, enhancements are more pronounced in E-commerce R\&S due to a higher proportion of search behaviors and item overlapping relative to recommendations.
Despite some methods like JSR and IV4REC+ showing degradation on certain metrics, ClardRec consistently demonstrates significant improvements, verified the transferring of item general features through counterfactual disentangling.

\begin{figure}[!t]
  \centering
  \includegraphics[width=0.75\linewidth]{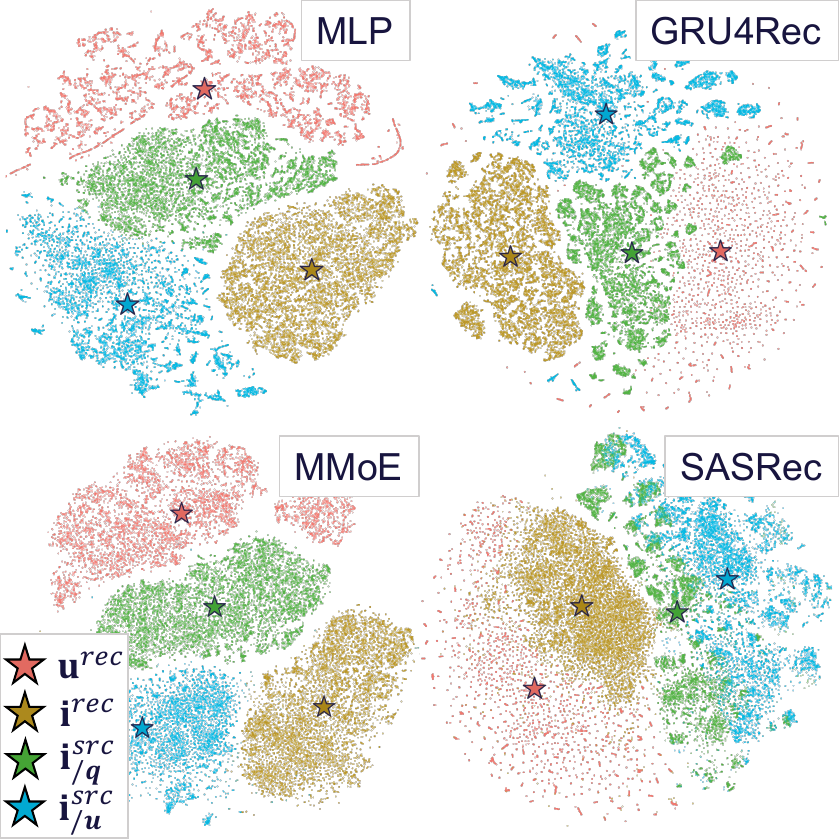}
  \caption{t-SNE visualization of query-independent and query-related item representations, user and item representations in recommendation using ClardRec across four backbones. We sampled 10,000 user-item click pairs for the collaborative filtering scenario and 1,000 user click sequences for the sequential scenario from the E-commerce R\&S dataset. Stars indicate the distribution centers of representations.}
 \label{fig:repr_visual}
 \vspace{-0.8em}
\end{figure}

\begin{figure}[!t]
  \centering
  \includegraphics[width=0.8\linewidth]{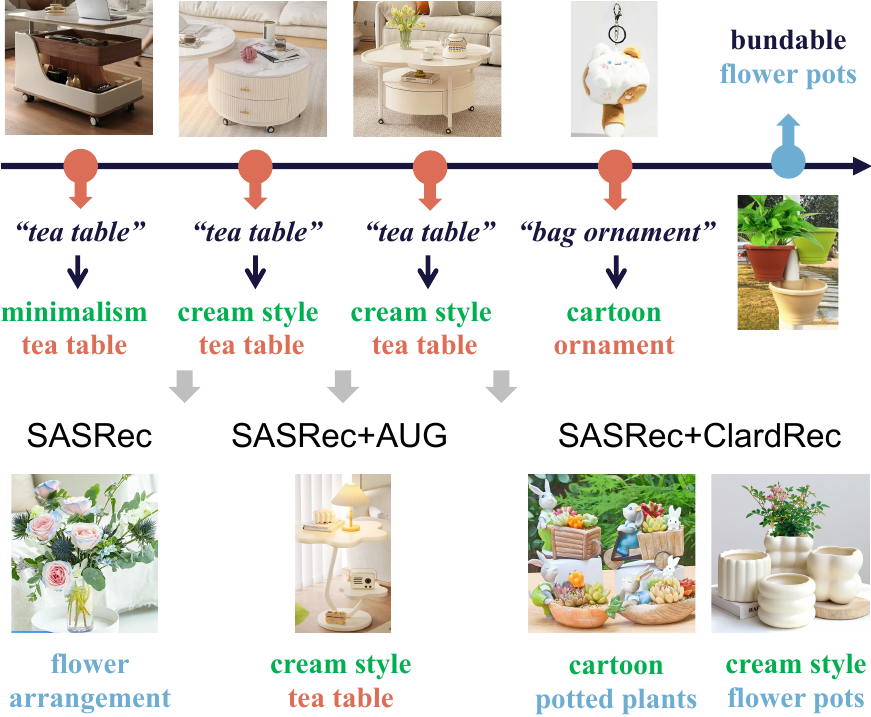}
  \caption{A real case of ClardRec enhancing recommendation with the SASRec backbone for a user in the E-commerce R\&S dataset. The orange words indicate the features in search. The blue words indicate the features in recommendation. The green words indicate the transferred item general features. }
 \label{fig:rec_case}
 \vspace{-1.2em}
\end{figure}

\subsection{Ablation Study (Q2)}

We conducted an ablation study to assess the impact of three key components of ClardRec.
``-CD'' removes counterfactual disentangling by replacing the query-independent item representation with the original item representation from search and omitting the corresponding loss.
``-FA'' removes feature augmentation by removing the gated fusion with query-independent item representation.
``-DA'' removes data augmentation by setting the corresponding loss term to zero.
As presented in Table~\ref{tab:abl_exp}, all three variants of ClardRec exhibit varying degrees of performance decline, underscoring their importance.
Notably, ``FA'' and ``CD'' show the most significant degradation, indicating that extracting and transferring query-independent item general features significantly enhances recommendation performance.
Additionally, ``DA'' also contributes to ClardRec's performance, showing drops across different backbones and baselines.
Its impact is less pronounced on some metrics for KuaiSAR.
This is because ``DA'' augments recommendation with the search data, compared with ``FA'' directly involving the candidate items in recommendation;
\textbf{the search data has an extremely small item overlapping ratio with recommendation in KuaiSAR (see Table~\ref{tab:data} for details}).

\subsection{Visualization of Disentangled Features (Q3)}

To verify whether ClardRec effectively decouples different representations of item features, Figure~\ref{fig:repr_visual} visualizes the two disentangled item representations, and the original user and item representations in recommendation.
According to our hypothesis, the query-independent item representations reflect user general interest and should therefore be closer to the user and item representations in recommendation.
As shown, the distribution aligns with our expectations, demonstrating that ClardRec successfully obtains distinct item representations and conducts efficient feature augmentation.

\subsection{Case Study of Feature Transferring (Q3)}

Figure~\ref{fig:rec_case} illustrates a real case of ClardRec enhancing recommendation for a user.
As shown, after querying ``tea table'' and ``ornament'' in search, the user clicked ``minimalism'' and ``cream style'' tea tables, and ``cartoon'' ornaments.
ClardRec excluded the query-related features and transferred such general features to recommendation.
After combining with the user interest ``flower pots'' in recommendation, ClardRec provided ``cartoon potted plants'' and ``cream style flower pots'', which satisfied the user's needs in recommendation.
In contrast, the original SASRec only captured the click behavior of ``flower pots'' and recommended ``flower arrangement''.
Augmented with search data, SASRec+AUG repeatedly recommended the interacted ``cream style tea table'' in search.
Neither method could explore the user's general interest across the two domains.

\section{Conclusion}
In this paper, we identify the issue that current search-enhanced recommendation methods directly incorporate search interactions into recommendation, which mixes user search-specific intents and thus leads to negative transfer effects.
To this end, we propose ClardRec which leverages search queries as anchors to disentangle query-independent item general features using a novel counterfactual disentangling approach.
Subsequently, our method employs feature and data augmentation modules to efficiently utilize these disentangled features.
Extensive experiments demonstrate that ClardRec achieves state-of-the-art performance compared to current mainstream search-enhanced recommendation methods.

\bibliographystyle{ACM-Reference-Format}
\bibliography{sample-base}

\appendix

\begin{table}[!t]
\centering
\caption{Data statistics of the preprocessed dataset in both collaborative filtering and sequential recommendation scenarios. ``src.'' and ``rec.'' are the abbreviations of search and recommendation. 
``attr.'' represents attributes. Compared with E-commerce R\&S, the KuaiSAR dataset has less item overlapping and search data proportion across the two domains.}
\label{tab:data}
\begin{tabular}{c|cc|cc}
\hline
Dataset         & \multicolumn{2}{c|}{KuaiSAR}       & \multicolumn{2}{c}{E-commerce R\&S}   \\ \hline
Scenario       & \multicolumn{1}{c|}{CF} & Sequential            & \multicolumn{1}{c|}{CF} & Sequential             \\ \hline
\#user in rec. & \multicolumn{1}{c|}{25.4k}   & \multirow{3}{*}{5.1k} & \multicolumn{1}{c|}{118.9k}  & \multirow{3}{*}{16.4k} \\
\#user in src.  & \multicolumn{1}{c|}{18.6k} &       & \multicolumn{1}{c|}{99.2k} &       \\
\#user in both  & \multicolumn{1}{c|}{18.2k} &       & \multicolumn{1}{c|}{40.0k} &       \\ \hline
\#item in rec.  & \multicolumn{1}{c|}{1.22m} & 0.10m & \multicolumn{1}{c|}{0.12m} & 0.36m \\
\#item in src.  & \multicolumn{1}{c|}{0.20m} & 0.11m & \multicolumn{1}{c|}{0.10m} & 0.35m \\
\#item in both  & \multicolumn{1}{c|}{0.01m} & 1.6k  & \multicolumn{1}{c|}{0.01m} & 56.5k \\ 
overlap ratio  & \multicolumn{1}{c|}{\textbf{1.0\%}} & \textbf{0.7\%}  & \multicolumn{1}{c|}{3.7\%} & 8.0\% \\ \hline
\#word          & \multicolumn{1}{c|}{53.9k} & 36.0k & \multicolumn{1}{c|}{51.4k} & 75.9k \\
\#click in rec. & \multicolumn{1}{c|}{3.73m} & 0.19m & \multicolumn{1}{c|}{1.29m} & 2.86m \\
\#click in src. & \multicolumn{1}{c|}{0.24m} & 0.13m & \multicolumn{1}{c|}{0.95m} & 2.53m \\ 
src. proportion & \multicolumn{1}{c|}{\textbf{6.1\%}} & 41.0\% & \multicolumn{1}{c|}{42.5\%} & 47.0\% \\ \hline
\#user attr.    & \multicolumn{2}{c|}{2}             & \multicolumn{2}{c}{2}              \\
\#item attr.    & \multicolumn{2}{c|}{4}             & \multicolumn{2}{c}{2}              \\ \hline
\end{tabular}
\vspace{-.5em}
\end{table}

\begin{figure*}[!t]
\centering
\includegraphics[width=\linewidth]{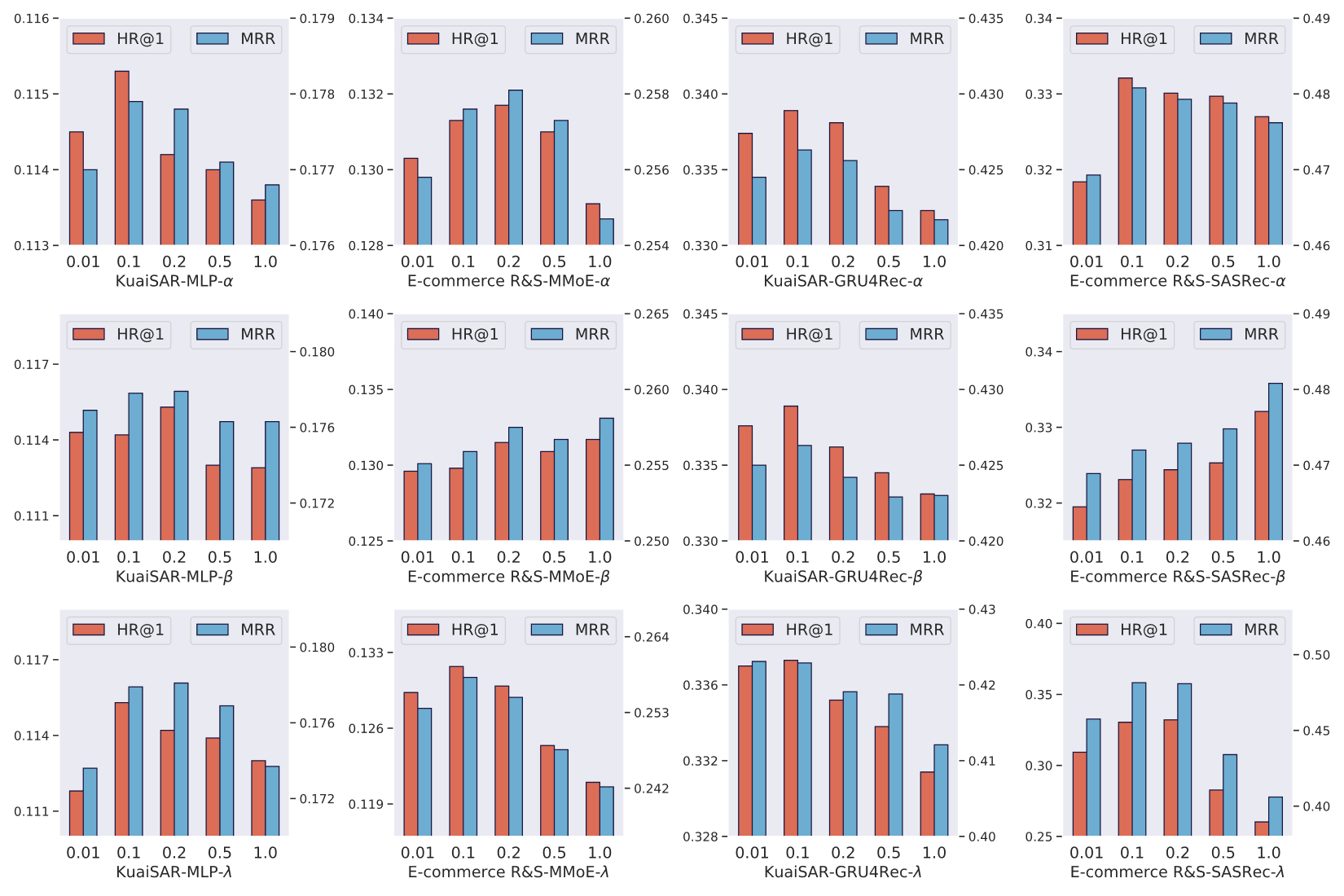}
\caption{
Hyper-parameter analysis to investigate the effects of counterfactual disentangling and data augmentation in ClardRec. We choose all the four backbones when they perform relatively best on the two datasets in both the collaborative filtering and sequential scenarios.
}
\label{fig:hyper}
\end{figure*}

\begin{table}[!t]
\centering
\caption{The hyper-parameter settings of ClardRec.}
\label{tab:hyper_setting}
\begin{tabular}{ccccccccc}
\hline
\multicolumn{9}{c}{KuaiSAR}                                                                                       \\ \hline
\multicolumn{1}{c|}{Backbone} & $lr$  & $bs$ & $L_b$ & $L_e$ & $\lambda$ & $\alpha$ & $\beta$ & $\gamma$ \\ \hline
\multicolumn{1}{c|}{MLP}      & 0.005 & 4096 & 3     & 3     & 0.1       & 0.1      & 0.2     & 0.5      \\
\multicolumn{1}{c|}{MMoE}     & 0.005 & 4096 & 3     & 3     & 0.2       & 1.0      & 0.1     & 0.1      \\
\multicolumn{1}{c|}{GRU4Rec}  & 0.001 & 64   & 1     & 3     & 0.02     & 0.1      & 0.1     & 0.1      \\
\multicolumn{1}{c|}{SASRec}   & 0.002 & 64   & 1     & 1     & 0.02     & 0.2      & 0.5     & 0.1      \\ \hline
\multicolumn{9}{c}{E-commerce R\&S}                                                                                 \\ \hline
\multicolumn{1}{c|}{Backbone} & $lr$  & $bs$ & $L_b$ & $L_e$ & $\lambda$ & $\alpha$ & $\beta$ & $\gamma$ \\ \hline
\multicolumn{1}{c|}{MLP}      & 0.005 & 1024 & 3     & 3     & 0.5       & 0.5      & 0.5     & 0.5      \\
\multicolumn{1}{c|}{MMoE}     & 0.005 & 1024 & 3     & 3     & 0.1       & 0.2      & 1.0     & 0.1      \\
\multicolumn{1}{c|}{GRU4Rec}  & 0.005 & 64   & 2     & 3     & 0.5       & 0.5      & 1.0     & 1.0      \\
\multicolumn{1}{c|}{SASRec}   & 0.005 & 64   & 1     & 1     & 0.2       & 0.1      & 1.0     & 1.0      \\ \hline
\end{tabular}
\end{table}


\section{Limitations and Future works}\label{app:limitation}

Our proposed ClardRec disentangles query-independent item features related to users' general interests to complement recommendations
However, there are some limitations that we could address in future research.
Currently, ClardRec requires overlapping users across the search and recommendation domains for general interest transfer.
We aim to overcome this limitation and adapt our method for more general scenarios without overlapping users.
Additionally, we will explore a more general search-enhanced recommendation framework that is not limited to ClardRec's current multi-encoder architecture.

\section{Experiment Details}

\subsection{Dataset Preprocessing}\label{app:dataset}

To accommodate both collaborative filtering and sequential recommendation scenarios, we preprocessed each dataset as follows:
(1) For the collaborative filtering scenario, each click behavior was stored as a data point, represented as a user-item pair for recommendation or a user-query-item triplet for search. Data points were split into training, validation, and test sets in an 8:1:1 ratio.
(2) For the sequential scenario, only the most recent 100 interactions for each user were retained.
Users with fewer than 10 search interactions were filtered out to ensure sufficient search data for enhancing recommendation.
Each user's most recent click behavior was used for testing, and the second most recent behavior was used for validation.
For negative sampling, 99 items were randomly selected that had never been clicked by the user (or under the query) for each interaction in both collaborative filtering and sequential scenarios.
Detailed statistics of the preprocessed datasets can be referred to in Table~\ref{tab:data}.

\subsection{Hyper-parameter Setting} \label{app:hyper_setting}
We provide the hyper-parameter settings of ClardRec in Table~\ref{tab:hyper_setting}. The notations in the table head indicate the learning rate, the batch size, the number of backbone layers, the number of item MLP encoder layers, and four loss balance factors in the total joint objective in Equation~\ref{eq:total_loss}.

\subsection{Backbone Details} \label{app:backbone}
We choose four user backbone encoders, MLP and MMoE for collaborative filtering, and GRU4Rec and SASRec for sequential recommendation.
We present the implementation details as follows.
\subsubsection{MLP}
This backbone~\cite{Huang2013} is directly designed as an $L$-layers MLP:
\begin{equation}
\textbf{u}^{dom} = \text{ReLU}\left(\cdots\text{ReLU}\left(u^{emb}\textbf{W}^{dom}_{1}+\textbf{b}^{dom}_{1}\right)\cdots\right)\textbf{W}^{dom}_{L}+\textbf{b}^{dom}_{L}
\label{eq:mlp_backbone}
\end{equation}
where $dom\in\{rec,src\}$ denotes the two versions in search and recommendation, respectively.
The first layer consists of a weight matrix $\textbf{W}^{dom}_{1}\in\real{d_u}{d_h}$ and a vector $\textbf{b}^{dom}_{1}\in\realx{d_h}$ to transform the user embedding into a $d_h$-dimension hidden vector.
The rest $L-1$ layers consist of linear weight matrices $\textbf{W}^{dom}_{l}\in\real{d_h}{d_h}$ and vectors $\textbf{b}^{dom}_{l}\in\realx{d_h}$ with $l\in\{2,\cdots,L\}$.

\subsubsection{MMoE}
This backbone~\cite{Ma2018} uses several MLPs as experts to derive different representations and combine them by learnable weights:
\begin{equation}
\textbf{u}^{dom} = \sum_{e=1}^{N}\omega_e\cdot\text{MLP}_e\left(\textbf{u}^{emb}\right)
\label{eq:mmoe_backbone}
\end{equation}
where $\omega_e$ is the $e$-th value of the weight vector $\boldsymbol{\omega}$ obtained by an another MLP with a softmax function:
\begin{equation}
\boldsymbol{\omega} = \text{softmax}\left(\text{MLP}_\omega\left(\textbf{u}^{emb}\right)\right).
\label{eq:weight_vector}
\end{equation}
In practice, we set the number of experts $N$ as 2.

\subsubsection{GRU4Rec}
This backbone~\cite{Hidasi2016} uses the gated recurrent unit network~\cite{cho2014} to sequentially encode the user's historical behaviors.
For the recommendation domain, we first obtain the sequence representation:
\begin{equation}
\textbf{u}^{rec}_{seq} = \text{GRU}_{rec}\left({\textbf{i}^{emb}_{rec,1},\cdots,\textbf{i}^{emb}_{rec,|\Hset^u_{rec}|}}\right)
\label{eq:gru}
\end{equation}
where $\textbf{i}^{emb}_{rec,j}$ is the embedding of the $j$-th item in the user click history $\Hset_{rec}^u$ in recommendation.
Then, this sequential representation is combined with the user embedding:
\begin{equation}
\textbf{u}^{rec} = \text{MLP}_{rec}\left([\textbf{u}^{emb}|\textbf{u}^{rec}_{seq}]\right)
\label{eq:gru_combine}
\end{equation}
For the search domain, we instead fuse the query representation in the sequence encoding as:
\begin{equation}
\textbf{u}^{src}_{seq} = \text{GRU}_{src}\left({[\textbf{i}^{emb}_{src,1}|\textbf{q}^{src}_{1}],\cdots,[\textbf{i}^{emb}_{src,|\Hset^u_{src}|}|\textbf{q}^{src}_{|\Hset^u_{src}|}]}\right)
\label{eq:gru_src}
\end{equation}
with a similar combination:
\begin{equation}
\textbf{u}^{src} = \text{MLP}_{src}\left([\textbf{u}^{emb}|\textbf{u}^{src}_{seq}]\right).
\label{eq:gru_combine_src}
\end{equation}
$\textbf{i}^{emb}_{src,j}$ and $\textbf{q}^{src}_{j}$ are the embeddings of the $j$-th item and query in the user click history $\Hset_{src}^u$ in search.

\subsubsection{SASRec}
This backbone~\cite{Kang2018} uses the self-attentive transformer encoders~\cite{Vaswani2017} to sequentially encode the user's historical behaviors.
It applies the similar encoding process as the backbone GRU4Rec but with a replaced self-attentive transformer encoder:
\begin{equation}
\begin{aligned}
&\textbf{u}^{rec}_{seq} = \text{SAE}_{rec}\left({\textbf{i}^{emb}_{rec,1},\cdots,\textbf{i}^{emb}_{rec,|\Hset^u_{rec}|}}\right),\\
&\textbf{u}^{src}_{seq} = \text{SAE}_{src}\left({[\textbf{i}^{emb}_{src,1}|\textbf{q}^{src}_{1}],\cdots,[\textbf{i}^{emb}_{src,|\Hset^u_{src}|}|\textbf{q}^{src}_{|\Hset^u_{src}|}]}\right).
\end{aligned}
\label{eq:sasrec}
\end{equation}

\begin{table}[!t]
\setlength{\tabcolsep}{3.9pt}

\caption{Ablation study of the confidence score in data augmentation on the E-commerce R\&S dataset. ``-CS'' denotes removing the confidence scores used in data augmentation..}
\label{tab:abl_confidence}
\begin{tabular}{c|l|cccc}
\hline
User encoder             & \multicolumn{1}{c|}{Method} & HR@1            & HR@5            & NDCG@5          & MRR             \\ \hline
\multirow{3}{*}{MLP}     & ClardRec                    & \textbf{0.1306} & \textbf{0.3613} & \textbf{0.2479} & \textbf{0.2541} \\
                         & -CS                         &   0.1280              &  0.3581               &    0.2452             &    0.2501             \\
                         & -FA                         & 0.1284          & 0.3598          & 0.2460          & 0.2520          \\ \hline
\multirow{3}{*}{MMoE}    & ClardRec                    & \textbf{0.1317} & \textbf{0.3710} & \textbf{0.2534} & \textbf{0.2581} \\
                         & -CS                         &   0.1305              & 0.3679                &  0.2511               & 0.2563                \\
                         & -FA                         & 0.1296          & 0.3655          & 0.2494          & 0.2551          \\ \hline
\multirow{3}{*}{GRU4Rec} & ClardRec                    & \textbf{0.2923} & \textbf{0.6199} & \textbf{0.4635} & \textbf{0.4437} \\
                         & -CS                         & 0.2882          & 0.6170          & 0.4601          & 0.4403          \\
                         & -FA                         & 0.2853          & 0.6101          & 0.4549          & 0.4395          \\ \hline
\multirow{3}{*}{SASRec}  & ClardRec                    & \textbf{0.3321} & \textbf{0.6561} & \textbf{0.5024} & \textbf{0.4808} \\
                         & -CS                         &   0.3284              &  0.6540               &    0.4967             &    0.4775             \\
                         & -FA                         & 0.3195          & 0.6444          & 0.4901          & 0.4689          \\ \hline
\end{tabular}
\end{table}

\section{Supplements for Experiments}\label{app:sup_exp}

\subsection{Hyper-parameter Analysis}

We conduct a hyper-parameter analysis to investigate the effects of loss balance factor $\alpha$, $\beta$, and $\lambda$, which respectively control the impact of counterfactual disentanglement, data augmentation, and the total auxiliary search task.
We vary the three hyper-parameters in the range of $\{0.01,0.1,0.2,0.5,1.0\}$ and present the results of ClardRec with those values in Figure~\ref{fig:hyper}.
As shown, all performance could peak in such ranges, which suggests that all the three learning objectives benefit ClardRec well.
Besides, the data augmentation shows more improvement when the corresponding loss factor $\beta$ becomes larger ($\beta=1.0$) in the E-commerce R\&S dataset rather than the lower values ($\beta=0.1,0.2$) in the KuaiSAR dataset.
This conforms to the discovery in the ablation study that a more similar distribution of items in the search and recommendation scenarios helps such data augmentation.
Furthermore, the effects of the counterfactual disentangling usually peak at a small value $\alpha = 0.1,0.2$ but decrease when the value gets larger, indicating that a larger value could dominate the learning of recommendation and cause degeneration.
As an auxiliary learning objective derived from search, the optimal values of $\lambda$ tend to fall around 0.1.
This range of values ensures an appropriate degree of knowledge transfer from search to recommendation. Larger values of $\lambda$ might lead the model to overly focus on search tasks, thus resulting in collapsed performance.

\subsection{Data Augmentation Analysis}

In the data augmentation component of ClardRec, we use $\hat{s}_{ui/q}$ as a confidence score and transform it as a loss function weight $\omega(\hat{s}_{ui/q})$ to augment recommendation.
This confidence score ensures the augmentation from search interactions containing more user general interest.
To verify this, we ablate the confidence scores by rewriting Equation~\ref{eq:da_pred} as
\begin{equation}
\mathcal{L}_{da}=CE(\hat{r}^{aug}_{ui},s_{uiq}),
\label{eq:da_pred_rewrite}
\end{equation}
and determine the performance shift.
To yield obvious results, we choose the E-commerce R\&S dataset which has a larger search interaction proportion and item overlapping.
As illustrated in Table~\ref{tab:abl_confidence}, the performance drop indicates that the confidence scores successfully filter more valuable search data for augmenting recommendation.

\end{document}